\documentstyle[amsmath,latexsym,twoside,fleqn,espcrc2]{article}

\title{Functional integration for Regge gravity}

\author{Pietro Menotti and Pier Paolo Peirano\address {Dipartimento di
Fisica dell'Universit\`a, Pisa 56100, Italy and INFN, Sezione di Pisa}%
        \thanks{This work is supported in part by M.U.R.S.T.}}
      
\begin{document}

\begin{abstract}
A summary is given of recent exact results concerning the functional
integration measure in Regge gravity.
\end{abstract}

\maketitle

\section{INTRODUCTION}

We shall describe an approach to the functional integration on
manifolds described by a finite number of parameters; the first part
will provide a general treatment while in the second part we shall
specialize to the $2$-dimensional Regge case were exact results can be
expressed in closed form. We wont delve in technical details, for
which we refer to the original papers \cite{pmppp1,pmppp2,pmppp3} but
we shall mainly concentrate on the general setting and on conceptual
issues.

In 1961 Regge \cite{regge} proposed a formulation of classical gravity
in which the continuous geometry is replaced by a geometry which is
piecewise flat, i.e. in which the curvature in concentrated on $D-2$
dimensional simplices being $D$ the dimension of spacetime. It is
remarkable that Einstein's equations derived by the discrete analogue
of the Einstein-Hilbert action assume a very simple form. The relation
of such a formulation to the classical continuum formulation, obtained
when the number of simplices goes to infinity has been thoroughly
studied in \cite{lee,cheeger}.  In classical gravity such an approach
can be used as an intrinsic geometric approximation scheme; at the
quantum level the Regge formulation has also been proposed as a way to
introduce a fundamental length into the theory \cite{lee}. Here we
shall understand the scheme at the quantum level as a mean to break
down the geometric degrees of freedom to a finite number, with the
idea that the continuum limit is obtained (or defined) when the number
of such degrees of freedom is allowed to go to infinity. Throughout
the treatment we shall refer to euclidean closed manifolds.

\section{GENERAL FORMULATION}

We shall consider in this section a general situation in which the
class of geometries described by a finite number of parameters is not
necessarily the Regge model. Diffeomorphisms play a key role in the
formulation of gravity and the viewpoint we shall adopt is to treat
them exactly at every stage. The class of geometries will be
parameterized by a finite number $M$ of invariants $l_i$ and described
by a gauge fixed metric $\bar{g}_{\mu\nu} (x,l)$. The functional
integration will be performed on the entire class of metrics
$[f^{\star}\bar{g}_{\mu\nu}(l)](x)$ with $f$ denoting the
diffeomorphisms \cite{jev}. The introduction of a metric for us is
crucial if we want to follow the analogy with gauge theory, being the
metric field $g_{\mu\nu}$ the analogue of the gauge field
$A_\mu$. The analogue of the gauge invariant metric of gauge theories
is the De Witt supermetric \cite{dewitt}
\begin{multline}
\label{dewittm}
(\delta g,\delta g) =  \\
\int \sqrt{g(x)}\, d^D x \,\delta
g_{\mu\nu}(x)
G^{\mu\nu\mu'\nu'}(x) \delta g_{\mu'\nu'}(x)
\end{multline}
where
\begin{multline}
G^{\mu\nu\mu'\nu'} = g^{\mu\mu'}g^{\nu\nu'} + g^{\mu\nu'}g^{\nu\mu'}-\\
\frac{2}{D}
g^{\mu\nu}g^{\mu'\nu'}+C g^{\mu\nu}g^{\mu'\nu'}
\end{multline} 
which is the most general
ultra--local distance in the space of the metrics, invariant under 
diffeomorphisms.
With regard to the reduction of the degrees of freedom we notice that
such a reduction will involve only the geometries not the
diffeomorphisms.
Since the integration on the latter is infinite dimensional the
related contribution will be a true functional integral (the
Faddeev--Popov determinant).
As usual when dealing with the differential structure of a manifold,
the charts and transition functions are to be given before imposing on
the differential manifold the metric structure; in other words if we
consider families of metrics on the same differential manifold the
transition functions have to be independent of the metric
itself. Such a feature is essential if we want that the variations
of the metric tensor appearing in the De Witt distance are to be
tensors under diffeomorphisms, or equivalently if the De Witt distance
has to be an invariant under diffeomorphisms.

In order to perform the integration on the metric we decompose
\cite{pmppp3} the general variation of $g_{\mu\nu}$ into two
orthogonal parts \begin{multline} \delta g_{\mu\nu} = [(F\xi)_{\mu\nu}
+ F(F^\dagger F)^{-1} F^\dagger \frac{\partial g_{\mu\nu}}{ \partial
l_i} \delta l_i ]+ \\ [1 - F (F^\dagger F)^{-1} F^\dagger]
\frac{\partial g_{\mu\nu}}{ \partial l_i} \delta l_i \end{multline}
where $\xi$ is the vector field representing the infinitesimal
diffeomorphism and $F^\dagger F$ is the Lichnerowicz-De Rahm operator
being $F$ defined by $(F\xi)_{\mu\nu} = \nabla_{\mu} \xi_{\nu} +
\nabla_{\nu}\xi_{\mu}$. It can be shown that the inverse of $F^\dagger
F$ is well defined from ${\rm Im} F^\dagger$ to ${\rm Im} F^\dagger$
\cite{pmppp3}. Obviously great simplifications would occur if the
gauge fixed metric could be chosen such as $\frac{\partial \bar
g_{\mu\nu}}{\partial l_i}\in \mbox{Ker}({F}^\dagger)$ however in
general such a choice cannot be accomplished \cite{pmppp3}. We recall
that given a distance, in our case the De Witt supermetric, it induces
a volume element on the tangent space of the metrics. This is the
direct generalization of how one computes the volume element in the
case of curved finite dimensional space.

A rather standard procedure allows now to factorize the infinite
volume of the diffeomorphism group and one reaches the integration
measure
\begin{equation}
\Pi_k dl_k \det (t^i, t^j)^{\frac{1}{2}} {\cal D}\mbox{et}
(F^\dagger F)^{\frac{1}{2}}
\label{see}
\end{equation}
with
\begin{equation}
t^i_{\mu\nu}=[1 - F (F^\dagger F)^{-1} F^\dagger]
\frac{\partial g_{\mu\nu}}{ \partial l_i}.
\end{equation}
In words, the first determinant represents the density of the
different geometries parameterized by the parameters $l_i$ while the
second determinant represents the gauge volume of such geometries (the
Faddeev-Popov determinant); it counts the number of ways different
metrics can be chosen to describe the same geometry. The source of this
second term is the fact that following the analogy with gauge theories
\cite{dewitt} we chose the metric as the fundamental variable.  Both
term are invariant not only under $l$-independent diffeomorphisms but
also under $l$-dependent diffeomorphisms \cite{pmppp3} and as such are
both true geometric invariants. In addition measure (\ref{see}) is
invariant in form under a change of the $M$ parameters which describe
the geometry; i.e. the result does not change whether to describe the
geometries we use a complete set of geodesic lengths, or a collection
of angles, areas etc. A property of the above measure is to be
dependent on the arbitrary constant $C$ which appears in the De Witt
metric. In fact both determinants in eq.(\ref{see})
depend on $C$ and one can show that the dependence in the general case
cannot cancel \cite{pmppp3}. This is not a surprise as such a constant
disappears only under integration over the conformal factor, to which
we shall now turn.

\subsection{Integration over the conformal factor}

We want to enlarge the treatment by replacing the integration
variables $l_{i}$ by a conformal factor $\sigma(x)$ \cite{mazur} and a
finite number of other parameters $\tau_{i}$ describing geometric
deformations transverse (i.e.\ non collinear) both to the
diffeomorphism and to the Weyl group.  

Thus the set of metrics we shall integrate on is given by
\begin{equation}
  g_{\mu\nu} (x,\tau,\sigma,f) = [f^{\star} e^{2\sigma}
\hat{g}_{\mu\nu}
  (\tau) ](x) \; .
\nonumber
\end{equation}
and now we have to compute the Jacobian $J(\sigma, \tau)$ such that
\begin{equation}
  {\cal D}[g] = J(\sigma, \tau) {\cal D}[f] {\cal D}[\sigma]
  \prod_{i} d\tau_{i}
\label{triangolo}
\end{equation}
being ${\cal D}[\sigma]$ the measure induced by the distance
\begin{equation}
(\delta \sigma, \delta \sigma)= \int \sqrt{g(x)} d^{D} x \delta
\sigma(x) \delta \sigma(x) \; .
\end{equation}
Proceeding as in the previous subsection the general variation of the
metric can be written as
\begin{multline}
\delta g_{\mu\nu} (x,\tau,\sigma,f) = (F \xi)_{\mu\nu}(x)
+ \\
2  [f^{\star} \delta \sigma \bar{g}_{\mu\nu}](x) + 
[f^{\star} \frac{\partial \bar{g}_{\mu\nu}}{\partial \tau_{i}}
  \delta \tau_{i}](x) \; ,
\end{multline}
with $\bar{g}_{\mu\nu}(x, \tau, \sigma) = e^{2\sigma} \hat{g}_{\mu\nu}
(x,\tau)$.  It is useful at this stage to introduce the traceless part
$P$ of $F$
\begin{equation}
(P\xi)_{\mu\nu} = (F\xi)_{\mu\nu} - \frac{g_{\mu\nu}}{D}
g^{\alpha \beta} (F \xi)_{\alpha \beta},
\end{equation}
the analogue $P^\dagger P$ of the Lichnerowicz-De Rahm operator and
the traceless tensor
\begin{equation}
k^{i}_{\mu\nu} = \frac{\partial g_{\mu\nu}}{\partial \tau_{i}} -
\frac{g_{\mu\nu}}{D} g^{\alpha\beta} \frac{\partial g_{\alpha
      \beta}}{\partial \tau_{i}}.
\end{equation}
The general variation of the metric can be written as
\cite{alvarez,moore,polch} 
\begin{multline}
\delta g_{\mu\nu} (x,\tau,\sigma,f) = (P\xi')_{\mu\nu} + f^{\star}
  2\delta \sigma' \bar{g}_{\mu\nu} (\tau,\sigma)  + \\
[1 - P (P^\dagger P)^{-1} P^\dagger] k^i_{\mu\nu} \delta
\tau_i
\end{multline}
where $\sigma'$ and $\xi'$ are proper translations of $\sigma$ and $\xi$.
The three terms are mutually orthogonal and exploiting the invariance
of the integrals under translations on the tangent space we have, apart
for a constant multiplicative factor
\begin{multline}
  J(\sigma, \tau) =  
{\cal D}\mbox{et} (\bar{P}^{\dag} \bar{P}
  )^{\frac{1}{2}} \times \\
    \left[ \det \left(  \bar{k}^{i}, ( 1 - \bar{P}
   (\bar{P}^{\dag} \bar{P})^{-1} \bar{P}^{\dag})
   \bar{k}^{j} \right) \right]^{\frac{1}{2}}. \\
\label{quadrato}
\end{multline}
The dependence on $f$ has disappeared due to the invariance of the De
Witt metric under diffeomorphisms and thus in eq.(\ref{triangolo}) the
infinite volume of the diffeomorphisms can be factorized away;
moreover in eq.(\ref{quadrato}) the dependence on $C$ has been
absorbed in an irrelevant multiplicative constant, as it happens in
two dimensions \cite{polyakov}. This is the result of having
integrated over all the conformal deformations. Again the first
determinant appearing in eq.(\ref{quadrato}) is a true functional
determinant while the second is an $M$-dimensional determinant.

We notice that, as we work with euclidean signature and on closed
manifolds i.e. compact manifolds without boundaries, the functional
determinants appearing in eq.(\ref{see}) and eq.(\ref{quadrato}) are
well defined through the usual $Z$-function regularization. This is
due to the fact that as it is easily checked both operators $F^\dagger
F$ and $P^\dagger P$ are elliptic for all $D\geq 2$. On the other hand
it is difficult for $D>2$ to extract the dependence of the two
determinants in (\ref{quadrato}) on the conformal factor $\sigma$. The
reason is that the usual procedure which works in two dimensions of
taking a variation with respect to $\sigma$ and then integrating back
stumbles into the appearance of the operator $P P^\dagger$ which in
$D>2$ is not elliptic and thus the usual heat kernel technology is no
longer available.

A finite dimensional approximation to eq.(\ref{quadrato}) is obtained
by restricting to a family of conformal factors parameterized by a
finite numbers of parameters $s = \{ s_{i} \}$.  Thus to the family
$f^{\star} e^{2\sigma(s)} \hat{g}_{\mu\nu} (\tau) $ it is associated
the measure
\begin{multline}
\displaystyle
\prod_{k} d\tau_{k} \prod_{i} ds_{i} \left[
\det(J^{\sigma}_{ij})\right]^{\frac{1}{2}} \times\\
\left[\det \left( k^{i}, ( 1 - P
  (P^{\dag} P)^{-1} P^{\dag})
  k^{j} \right) 
{\cal D} \mbox{et} (P^{\dag} P ) \right]^{\frac{1}{2}}\\
\label{measure2}
\end{multline}
where $J^{\sigma}_{ij} = \int d^{D} x \sqrt{\hat{g}}\;  e^{D\sigma}
\frac{\partial \sigma}{\partial s_{i}} \frac{\partial \sigma}{\partial
s_{j}} $.

We now turn to the $D=2$ case
where explicit results can be obtained.
  
\section{THE D=2 CASE:  TWO DIMENSIONAL REGGE GRAVITY IN THE
CONFORMAL GAUGE}

We need now to specialize eq.(\ref{measure2}) to the two dimensional
case. Here the role of the $\tau_i$ is played by the Teichm\"uller
parameters which are absent for genus $0$, are two in number for genus
$1$ and $6h-6$ for higher genus; in addition, contrary to what happens
in higher dimensions were the generic geometry has no conformal
Killing vector field, in two dimensions every topology carries its own
conformal Killing vector fields which are 6 for genus $0$ (sphere
topology), 2 for genus $1$ (torus topology) and are simply absent for
higher genus. In presence of conformal Killing vectors
eq.(\ref{measure2}) goes over to \cite{alvarez,moore,polch}
\begin{multline}
{\cal{D}} [\sigma] \,  \frac{d\tau_{i}}{v(\tau)}
\sqrt{\frac{{\det}'(P^{\dag}P)}{\det(\phi_{a}, \phi_{b})
    \det(\psi_{k},\psi_{l})}}
\displaystyle \times  \\
\det (\frac{\partial
g}{\partial\tau_{n}},\psi_{m}).
\label{part}
\end{multline}  
$\cal{D}[\sigma]$ is the functional
integration measure induced by the metric
\begin{equation}
(\delta \sigma^{(1)}, \delta \sigma^{(2)}) =  \int \sqrt{\hat{g}}~
e^{2\sigma} \delta \sigma^{(1)} \delta \sigma^{(2)}.
\label{quindici}
\end{equation}
$\phi_{a}$ and $\psi_{k}$ are respectively the zero modes of $P$ and
$P^{\dag}$; $v(\tau)$ represents the volume of the conformal
transformations.  ${\rm det}'(P^\dagger P)$ stays for the determinant
from which the zero modes have been excluded. It is well known that
$P$ acts diagonally on the column vector $(\xi_{\bar{\omega}} ,
\xi_{\omega})$ by transforming it into $(h_{\bar{\omega}
\bar{\omega}}, h_{\omega\omega})$ which represent a traceless
symmetric tensor in two dimensions. In the conformal gauge the
operator $L$ which takes $\xi_{\bar{\omega}}$ into $h_{\bar{\omega}
\bar{\omega}}$ and its adjoint are given by
\begin{multline}
    L = e^{2\sigma} \frac{\partial}{\partial \bar\omega}
  e^{-2\sigma} \quad \mbox{and} \quad
  L^{\dag} = - e^{-2\sigma} \frac{\partial}{\partial \omega}
\\ \label{opland}
\end{multline}
and we have $\det'(P^{\dag}P) = [\det'(L^{\dag}L)]^{2}$.

In $D=2$ the singularities of the Regge geometry are confined to
isolated points where conical defects (positive or negative) are
present. Our problem therefore will be to compute the determinant
appearing in eq.(\ref{part}) on a two dimensional surface which is
everywhere flat except for isolated conical singularities. The
determinant of $L^\dagger L$ will be defined through the $Z$-function
technique where
\begin{equation}
  Z_{K}(s) = \frac{1}{\Gamma (s)} \int_{0}^{\infty} \! dt \: t^{s-1}
\mbox{Tr}'(e^{-tL^{\dag}L})
\end{equation}
and 
\begin{multline}
-\log({\det}'(L^{\dag}L))= \dot{Z}_{K}(0) = \gamma_E Z_{K}(0) + \\
{\mbox{Finite}}_{\epsilon \rightarrow 0}
\int^\infty_\epsilon \frac{dt}{t} \mbox{Tr}'(e^{-tL^{\dag}L}) \, .
\end{multline}
The standard procedure is to compute the change of $\dot{Z}_{K}(0)$
under a variation of the conformal factor
\begin{multline}
- \delta \log \left[
  \frac{\det'(L^\dagger L)}{
    \det(\Phi_a,\Phi_b)\det(\Psi_l,\Psi_m)}
\right]=  \\
 \gamma_{E} \delta c_{0}^{K} +
{\mbox{Finite}}_{\epsilon \rightarrow  0}
\; \mbox{Tr}\:\left[ 4\delta\sigma{\cal K}(\epsilon)
-2\delta\sigma {\cal H}(\epsilon) \right], \\
\label{covva}
\end{multline}
and then integrating back the result.
 In the previous equation
$K=L^{\dag}L$, $H=LL^{\dag}$, ${\cal K}$ is the heat kernel of $K$ and
${\cal H}$ is the heat kernel of $H$; $c_{0}^{K}$ is the constant term
in the asymptotic expansion of the trace of the heat kernel ${\cal
K}(t)$ and is related to $Z_{K}(0)$ by
\begin{equation}
c_{0}^{K} = Z_{K}(0) + \mbox{dim}(\mbox{Ker }K).
\end{equation}
$\Phi_{a}$ and $\Psi_{i}$ are the zero modes of $K$ and $H$
respectively.  The central point in the evaluation of the r.h.s.\ of
eq.(\ref{covva}) will be the knowledge of $c_{0}^{K}$ and of ${\cal
K}(t)$ and ${\cal H}(t)$ on the Regge manifold for small $t$. As is
well known such quantities are local in nature and thus we shall start
by computing them on a single cone.

\subsection{ Self-adjoint extension of the \\Lichnerowicz-De Rahm operator}

In order to compute the small time behavior of the heat kernel around a
conical singularity we need to solve the eigenvalue equation
$(L^\dagger L)\xi = \lambda \xi$ on a cone. This is done as usual by
decomposing $\xi$ in circular harmonics. A peculiar aspect of the
problem is that for a generic opening $\alpha$ of the cone ($\alpha=1$
is the plane) there is a number of partial waves for which both
solutions (the ``regular'' and the ``irregular'' at the origin) are
square integrable in the invariant metric; as well known such a
circumstance poses the problem of the correct self-adjoint extension of
the Lichnerowicz-De Rahm operator. Originally \cite{pmppp1} the problem was
solved by regularizing the tip of the cone by a segment of sphere or
Poincar\'e pseudosphere and then taking the regulator to zero. A more
general approach was given subsequently in \cite{pmppp2} by considering all
self-adjoint extension of $L^\dagger L$ and imposing on them the
restrictions due to the Riemann-Roch relation
\begin{multline}
  2( c_{0}^{K} - c_{0}^{H} ) = \mbox{dim }(\mbox{ker }(P^\dagger P))
- \\ 
\mbox{dim }(\mbox{ker }(PP^\dagger)) = 3\chi
\end{multline}
being $\chi=2 -2 h$ the Euler characteristic of the surface of genus $h$.
The two procedures give exactly the same results; in particular 
\begin{equation}
  \label{ckris}
  c_{0}^{K} = \frac{1 - \alpha^{2}}{12\alpha} + \frac{(\alpha -
    1)(\alpha - 2 )}{2\alpha}.
\end{equation}
and
\begin{equation}
  \label{chris}
  c_{0}^{H} = \frac{1 - \alpha^{2}}{12\alpha} + \frac{(2\alpha -
    1)(2\alpha - 2 )}{2\alpha}.
\end{equation}

A posteriori the agreement between the two methods is not surprising 
since the Riemann-Roch relation is a statement related to 
the topology of the surface which is correctly provided by the
smoothing process. In different words the boundary conditions imposed by
the Riemann-Roch theorem are those which reflect the compactness of
the manifold.
Outside the interval $\frac{1}{2} < \alpha < 2$ is not possible to
satisfy the Riemann--Roch relation within the realm of
$L^2$-functions.

We now turn to the explicit computation of the determinants referring
to the simpler cases of $h$ equal $0$ and $1$.

\subsection {Sphere topology}

The conformal factor describing a Regge geometry with the topology of the
sphere is given by \cite{foer,aurell,pmppp1}
\begin{equation}
e^{2\sigma} = e^{2\lambda_{0}} \prod_{i=1}^{N} | \omega - \omega_{i}
|^{2(\alpha_{i} -1)}
\label{fattore}
\end{equation}
with $0<\alpha_{i}$ and $\sum_{i=1}^{N} ( 1 - \alpha_{i} ) = 2$.
As it happens on the continuum such a conformal factor is unique
\cite{ginsparg} up to the 6 parameter $SL(2,C)$ transformations 
corresponding to the six conformal Killing vectors of the sphere
\begin{eqnarray}
\label{conftrans}
\displaystyle \omega_{i}' = \frac{a\omega_{i} + b}{c\omega_{i} + d },
\qquad \alpha_{i}' = \alpha_{i} \\
 \lambda_{0}' = \lambda_{0} + \sum_{i=1}^{N} (\alpha_{i} -1 ) \log
|\omega_{i} c + d| , \nonumber 
\end{eqnarray}
with the complex parameters satisfying $ad -bc = 1$. 
Under the written transformation the conformal factor
\begin{multline}
\sigma \equiv \sigma(\omega ; \lambda_{0},
\omega_{i}, \alpha_{i} )= \\
\lambda_0+\sum_i(\alpha_i-1)\log|\omega-\omega_i|
\label{ventisei}
\end{multline}
goes over to
\begin{equation}
\sigma'(\omega'; \lambda_{0}, \omega_{i}, \alpha_{i}) =
\sigma(\omega';  \lambda_{0}', \omega_{i}', \alpha_{i}')
\end{equation}
where $\omega'_i$, $\alpha'_i$ and $\lambda_{0}'$ are given by
eq.(\ref{conftrans}).  The area $A$
\begin{equation}
A=e^{2\lambda_0}\int d^2\omega |\omega-\omega_i|^{2(\alpha_i-1)}
\end{equation}
being a geometric invariant is left unchanged.  It is a remarkable
feature of the family of Regge conformal factors to be closed under
such $SL(2,C)$ transformations.  Such a description is equivalent to
usual one in terms of link lengths; in fact from the Euler relation
$F+V = H+2$ with $H=\frac{3}{2} F$ we get $H=3V -6$, where $-6$
corresponds to the $6$ conformal Killing vectors of the sphere.

Substituting now the variation of the conformal factor
eq.(\ref{ventisei}) into eq.(\ref{covva}) and integrating back the
result, one obtains \cite{pmppp1,pmppp2}
\begin{multline}
\label{primoris}
  \log \sqrt{\frac{{\det}'(P^{\dag}P)}{\det(\phi_{a}, \phi_{b})}} \\ 
  \displaystyle =\frac{26}{12} [\sum_{i,j\neq i}
    \frac{(1-\alpha_{i})(1-\alpha_{j})}{\alpha_{i}} \log |\omega_{i} -
    \omega_{j}| \\
+ \lambda_{0} \sum_{i} (\alpha_{i} - \frac{1}{\alpha_{i}})
    - \sum_{i} F(\alpha_{i})] \; .
\end{multline}
$F$ is a smooth function for which an integral representation can be
given \cite{pmppp2}. By direct substitution one verifies that 
eq.(\ref{primoris}) is
invariant under the $SL(2,C)$ transformation eq.(\ref{conftrans}). We
remark that the rational structure in the $\alpha_i$ appearing in
eq.(\ref{primoris}) is essential for such invariance to be
accomplished.

We notice that apart from the term $\sum_{i=1}^{N} F(\alpha_{i})$
eq.(\ref{primoris}) is exactly $-26$ times the conformal anomaly for
the
scalar field as computed by Aurell and Salomonson \cite{aurell}.  In the
continuum limit $N \rightarrow \infty$, the $\omega_{i}$ become dense
and the $\alpha_{i} \rightarrow 1$, always with $\sum_{i=1}^{N}
(1-\alpha_{i})=2$. In such a limit $\sum_{1=1}^{N} F(\alpha_{i})$ goes
over to the topological invariant $N\, F(1) - \chi F'(1)$, while the
remainder goes over to the well known continuum expression. In fact
we have
\begin{equation}
\frac{1}{2\pi} \log | \omega - \omega' | = \frac{1}{\Box} (\omega,
\omega')
\end{equation}
and for any region $V$ of the plane $\omega$
\begin{multline}
\int_V d^2\omega e^{2\sigma} R = -2 \int_{V} d^2\omega
\Box \sigma \\ = 4\pi \sum_{i: \omega_{i} \in V} (1 -\alpha_{i}).
\end{multline}
Thus the r.h.s. of eq.(\ref{primoris}) goes over to
\begin{multline}
\displaystyle
\label{contliouv}
\frac{26}{96\pi}[ \int  d^{2}\omega \, d^{2}\omega' \;(\sqrt{g}
  R)_{\omega }
\frac{1}{\Box}(\omega, \omega')(\sqrt{g} R)_{\omega'} \\ 
- 2 (\log \frac{A}{A_0})
\int \: d^{2} \omega\,\sqrt{g} R] 
\end{multline}
where $A_0$ is the value of the area for $\lambda_0=0$;
eq.(\ref{contliouv}) is the correct continuum result.

We notice that as it happens on the continuum, the F.P. contribution
eq.(\ref{primoris}) is not the result of integrating on the
fluctuations of the geometry but of integrating on the
diffeomorphisms, while keeping the geometry (in our case described by
the conformal factor) exactly fixed.  One should not confuse the
diffeomorphisms with the zero modes of the action i.e. changes in the
geometry which leave the action invariant.  

On the numerical front
accurate simulation have been given of two dimensional gravity, both
pure and coupled with Ising spins, by adopting the measure $\prod_i
dl_i/l_i$.  The results are consistent with the Onsager exponents
and in definite disagreement with the KPZ exponents \cite{holm} while
the situation for the string susceptibility is still unclear
\cite{holm,bock}. That measures of type $\prod_i dl_{i} f(l_{i})$ fail
to reproduce the Liouville action can be understood by the following
argument \cite{pmppp2}: on the continuum for geometries which deviate
slightly from the flat space one can compute approximately the
Liouville action by means of a one loop calculation. If one tries to
repeat a similar calculation for the Regge model with the measure
$\prod_i dl_{i} f(l_{i})$ one realizes that being the Einstein action
in two dimensions a constant, the only dynamical content of the theory
is played by the triangular inequalities. But at the perturbative
level triangular inequalities do not play any role and thus one is
left with a factorized product of independent differentials which
bears no dynamics and thus no Liouville action.

\subsection{Integration measure for the conformal factor}

It is not enough to give the Faddeev- Popov determinant
eq.(\ref{primoris}); one must also give the explicit form of the
measure ${\cal D}[\sigma]$ in the Regge case.  According to
eq.(\ref{quindici}) the distance between two nearby configurations
$\sigma$ and $\sigma + \delta \sigma$ of the conformal factor is given
by
\begin{equation}
  (\delta \sigma, \delta \sigma) = \int \! d^{2}\omega \, e^{2\sigma}
\, \delta \sigma \, \delta \sigma \, .
\label{confmetr}
\end{equation}
Such an expression is a direct outcome of the original De-Witt
measure (\ref{dewittm}). From eq.(\ref{confmetr}) it follows that having
parameterized the
Regge surface by means of the $3N$ variables $p_{i}$
\begin{multline}
  \{ p_{1},\ldots, p_{3N}\} \equiv \{ \omega_{1,x}, \omega_{1,y},
  \omega_{2,x}, \omega_{2,y}, \ldots, \\
\omega_{N,x}, \omega_{N,y},
  \lambda_{0}, \alpha_{1}, \alpha_{2}, \ldots, \alpha_{N-1} \}
\end{multline}
${\cal D}[\sigma]$ is given by
\begin{equation}
  {\cal D}[\sigma] = \sqrt{\det J} \prod_{k=1}^{N} d^{2}\omega_{k} \;
  \prod_{i=1}^{N-1} d\alpha_{i} d\lambda_{0} \, 
\end{equation}
being $J$ the $3N \times 3N$ matrix
\begin{equation}
  \label{jac}
  J_{ij} = \int d^{2}\omega \, e^{2\sigma} \, \frac{\partial
    \sigma}{\partial p_{i} } \frac{\partial \sigma}{\partial p_{j}},
\end{equation}
with $\alpha_{N} = \sum_{i=1}^{N-1} (1 -\alpha_{i}) -1$.

We notice that all $J_{ij}$ are given by convergent integrals except
those involving two $\omega_{i}$ with the same indexes, which
converge only for $\alpha_{i} > 1 $. For example we
have
\begin{equation}
  \label{diagdiv}
  J_{\omega_{i,x}\omega_{i,x}} = (\alpha_{i} - 1 )^{2} \int d^{2}
\omega
  \, e^{2\sigma} \: \frac{(\omega_{i,x} -\omega_{x})^{2}}{ |\omega -
    \omega_{i}|^{4}}
\end{equation} 

For $\alpha_i -1 \rightarrow 0$ $J_{ij}$ vanishes and for $\alpha_i <1$
it has a well defined analytic continuation \cite{pmppp2}. As a result of the
factor $\delta_i = 1- \alpha_i$ appearing in front of all rows of the type
$J_{\omega_{i,x}, p_{j}}$, $\det J$ vanishes whenever an $\alpha_{i}$
equals $1$, as expected from the fact that in such a situation the
position of the vertex $i$ is irrelevant in determining the metric.
A remarkable property of ${\cal D}[\sigma]$ is to be invariant under
the $SL(2,C)$ group with the result that the whole theory is invariant
under such transformations. The measure can also be written as
\begin{equation}
  \label{jform}
\prod_{l=1}^{3N} dp_l  e^{3N\lambda_{0}}  \prod_{k=1}^{N} |\alpha_i-1|
\prod_{i, j>i} |\omega_{i} -  \omega_{j}|^{4 \beta_{ij}} Y
\end{equation}
where
\begin{equation}
  \label{betasol}
  \beta_{ij} = \frac{3}{2} \frac {N}{N-2} (\frac{2}{N-1}
  -\delta_{i} -\delta_{j}) - \frac{2}{N-1}
\end{equation}
and $Y$ is a function only of the $\alpha_i$ and the harmonic ratios
of the $\omega_{i}$.

The invariance under the finite dimensional group $SL(2,C)$ is
sufficient to prove that the field $\exp(2q\sigma(x))$ maintains its
canonical dimension $q$ which is in agreement with the analysis of the
continuum theory in presence of the Weyl covariant measure \cite{hoker}.

\subsection{Torus topology}

The most general metric, modulo diffeomorphisms, is given by a flat
metric $\hat{g}_{\mu \nu}(\tau_{1}, \tau_{2})$ times a conformal
factor $e^{2\sigma}$.  $\tau_{1}$ and $\tau_{2}$ are the two
Teichm\"{u}ller parameters in terms of which, with $\tau=\tau_{1}+i
\tau_{2}$,
\begin{equation}
ds^{2}= dx^{2}+2\tau_{1}dx dy+ |\tau|^{2} dy^{2}
\end{equation}
and the fundamental region has been taken the square $ 0 \leq x <1, 0
\leq y<1$. The conformal factor for the torus can be expressed in terms
of the torus Green's function
\begin{multline}
G(\omega - \omega' | \tau)  = \frac{1}{2\pi} \log \left|
    \frac{\vartheta_{1} (\omega -
\omega'| \tau )}{\eta(\tau)} \right| \\ - \frac{(\omega_{y} -
\omega_{y}')^{2}}{2 \tau_{2}}
\end{multline}
being $\vartheta_{1}(\omega | \tau)$ the Jacobi $\vartheta$--function
and
\begin{equation}
\displaystyle
\eta(\tau)=  e^{\frac{i\pi\tau}{12}} \prod_{n=1}^{\infty} [ 1 -
e^{2in\pi\tau} ] .
\end{equation}
{}From the relation 
\begin{equation}
R(e^{2\sigma}\hat{g})=e^{-2\sigma}(R(\hat{g})-2 \hat{\Box}\sigma)
\end{equation} 
we have that the conformal factor for the Regge surface with the
topology of the torus is given by
\begin{multline}
\label{sigmatoro}
\sigma (\omega) = \lambda_{0} + \sum_{i=1}^{N} (\alpha_{i} -1) \left[
\log \left|
    \frac{\vartheta_{1} (\omega - \omega_{i}| \tau )}{\eta(\tau)}
\right| \right.\\
\left.  - \frac{\pi}{\tau_{2}}(\omega_{y} -\omega_{i,y})^{2} \right].
\end{multline}
Thus the physical degrees of freedom are $3N$: in fact in addition to
the $2N$ $x_i,\, y_i$ we have $N-1$ independent angular deficits
($\sum_{i=1}^{N} (\alpha_{i} -1) = 0$), two Teichm\"{u}ller parameters
and $\lambda_{0}$, to which we must subtract the two conformal Killing
vectors of the torus. We have the same number of physical degrees of
freedom as the number of bones in a Regge triangulation of the torus
with $N$ vertices as it can be easily checked through the Euler
relation for a torus ($F + V = H = 3F/2$, from which $H=3V$).
The derivation of the Liouville action proceeds similarly as for the
sphere topology with the final result for the partition function
\begin{equation}
\label{toruspart}
\displaystyle
\int {\cal D}[\sigma] \frac{d^{2}\tau}{\tau_2}|\eta(\tau)|^4
\displaystyle{e^{\frac{26}{12} S_{l}} }
\end{equation}
where
\begin{multline}
\label{torusliouv}
S_{l}=\sum_{i,j\neq i} \frac{(1 - \alpha_{i})(1
    -\alpha_{j})}{\alpha_{i}} \left[  \log \left| \frac{\vartheta_{1}
    (\omega_{j} - \omega_{i}| \tau )}{\eta(\tau)} \right| \right. \\
- \left. \frac{\pi}{\tau_{2}} (\omega_{i,y} - \omega_{j,y})^{2} \right]  
 + (\lambda_{0} - \log | 2\pi \eta^{2} | ) \times \\
\sum_{i} (\alpha_{i} -
    \frac{1}{\alpha_{i}}) - \sum_{i} F(\alpha_{i}).
\end{multline}

In the continuous limit eq.(\ref{torusliouv}) goes over to the well
known
expression
\begin{equation}
\displaystyle
\label{contliouvtorus}
\frac{1}{8\pi} \int  d^{2}\omega \, d^{2}\omega' \;(\sqrt{g}
  R)_{\omega }
\frac{1}{\Box}(\omega, \omega')(\sqrt{g} R)_{\omega'}
\end{equation}

\subsection{Modular invariance}

It is possible now to give an explicit, non formal proof of the
modular invariance of the theory.  In eq.(\ref{toruspart}) it is well
know that $d^2\tau |\eta(\tau)|^4/\tau_2$ is invariant under the
modular transformation
\begin{equation}
\label{transmod}
\tau \longrightarrow \tau' = \frac{\tau a + b}{\tau c + d }
\end{equation}
with $(a,b,c,d) \in {\bf Z}$ and $ad -bc =1$.  Thus we are left to
prove the modular invariance of $\int {\cal D}[\sigma]
e^{\frac{26}{12} S_{l}}$.

This is achieved by accompanying the change in $\tau$ by a proper
change in the integration variables $\omega_{i}, \lambda_{0}$ given by
\begin{equation}
\label{modtrans}
\omega' = \frac{\omega}{\tau c + d}
 \qquad \lambda_{0}' = \lambda_{0} + \log |
  \tau c + d |
\end{equation}
where the transformation of $\lambda_0$ follows from the transformation
of $\sigma$ and the modular invariance of $G$, i.e. $G(\omega
-\omega_{i} | \tau) = G(\omega' -\omega_{i}' | \tau')$. 
$S_{l}$, as given by eq.(\ref{torusliouv}), is invariant under
transformations
(\ref{modtrans}), (\ref{transmod}) because of the just cited modular
invariance of the Green function and because
\begin{equation}
\eta \left (\frac{a\tau + b}{ c\tau + d} \right) =
e^{i\phi}(c\tau + d)^{\frac{1}{2}} \eta(\tau)
\end{equation}
compensates the change in $\lambda_{0}$. 

Again the invariance of the
area leaves $\sqrt{J} \prod_{i=1}^{N} d^{2}\omega_{i}$$ d\lambda_{0}
\prod_{j=1}^{N-1} d\alpha_{j}$ invariant and this
concludes the proof of modular invariance.

\section{CONCLUSIONS}

Starting from the De Witt distance among metrics we have derived, when
we restrict ourselves to a subclass of geometries described by a
finite number of parameters, the ensuing expression for the functional
integration measure. The results eq.(\ref{see}) and
eq.(\ref{measure2}) follow directly from the De Witt supermetric
without any other additional input. They are mathematically well
defined provided we work in the euclidean and on closed manifolds,
i.e. compact manifolds without boundaries. In two dimensions we can
give an exact expression for the F.P. both for the sphere
and torus topology. Higher genus require the knowledge of the Green
function on a surface of constant negative curvature and given
Teichm\"uller parameters.  The integration measure on the conformal
factor appears in the form of a $3N\times 3N$ determinant which has
the correct invariance properties as the F.P. term. The matrix
elements of the determinant are given in terms of homogeneous integrals of
dimension $\omega^{-2}$ of the type which appeared in the old
conformal field theory. It would be of interest the closed
evaluation of the determinant at least in some simple example
or a rigorous estimate of the determinant for large $N$,
which is of importance for the continuum limit.

The exact analytical evaluation of the functional determinants
appearing in eq.(\ref{see}) and eq.(\ref{measure2}) in dimension
higher that two has not yet been performed and we have discussed the
main technical differences with respect to two dimensions. For $D>2$ the
extraction of the $\sigma$- dependence of the ${\cal D}{\rm et}$
appearing in eq.(\ref{measure2}) is an old standing problem (see
e.g. \cite{deser}).  In the meantime the simpler approach with the
measure (\ref{see}) appears more viable in $D>2$. According to the
treatment of section (2.1), here one expects that for $M\rightarrow
\infty$ the dependence on the parameter $C$ should disappear; actually
such a dependence could be taken as a measure of the approach to the
continuum limit. The $D=3$ which appears simpler than the $D=4$ case
bears some relation with recent result of classical $2+1$
dimensional gravity in presence of point particle and progress in that
field may also be helpful \cite{ciafaloni,welling}.

\end{document}